# Puddle formation, persistent gaps, and non-mean-field breakdown of superconductivity in overdoped $(Pb,Bi)_2Sr_2CuO_{6+\delta}$

Willem O. Tromp[1*], Tjerk Benschop[1*], Jian-Feng Ge[1], Irene Battisti[1], Koen M. Bastiaans[1,2], Damianos Chatzopoulos[1], Amber Vervloet[1], Steef Smit[3], Erik van Heumen[3,4], Mark S. Golden[3], Yinkai Huang[3], Takeshi Kondo[5], Yi Yin[6,7], Jennifer E. Hoffman[8], Miguel Antonio Sulangi[9], Jan Zaanen[10], Milan P. Allan[1†]

1. Leiden Institute of Physics, Leiden University, 2333 CA Leiden, The Netherlands
2. Department of Quantum Nanoscience, Kavli Institute of Nanoscience, Delft University of Technology, 2628 CJ Delft, The Netherlands
3. Institute of Physics, University of Amsterdam, 1098 XH Amsterdam, The Netherlands
4. QuSoft, 1098 XG Amsterdam, The Netherlands
5. Institute for Solid State Physics, University of Tokyo, Kashiwa, Japan
6. Zhejiang Province Key Laboratory of Quantum Technology and Device, Department of Physics, Zhejiang University, Hangzhou 310058, China
7. Collaborative Innovation Centre of Advanced Microstructures, Nanjing University, Nanjiang 210093, China
8. Department of Physics, Harvard University, Cambridge, Massachusetts 02138, USA
9. Department of Physics, University of Florida, Gainesville, Florida 32611, USA
10. Institute-Lorentz for Theoretical Physics, Leiden University, 2333 CA Leiden, The Netherlands

* These authors contributed equally.
† Corresponding author: milan.allan@gmail.com

**The cuprate high-temperature superconductors exhibit many unexplained electronic phases, but it was often thought that the superconductivity at sufficiently high doping is governed by conventional mean-field Bardeen-Cooper-Schrieffer (BCS) theory[1]. However, recent measurements show that the number of paired electrons (the superfluid density) vanishes when the transition temperature $T_c$ goes to zero[2], in contradiction to expectation from BCS theory. The origin of this anomalous vanishing is unknown. Our scanning tunneling spectroscopy measurements in the overdoped regime of the $(Pb,Bi)_2Sr_2CuO_{6+\delta}$ high-temperature superconductor show that it is due to the emergence of puddled superconductivity, featuring nanoscale superconducting islands in a metallic matrix[3,4]. Our measurements further reveal that this puddling is driven by gap filling, while the gap itself persists beyond the breakdown of superconductivity. The important implication is that it is not a diminishing pairing interaction that causes the breakdown of superconductivity. Unexpectedly, the measured gap-to-filling correlation also reveals that pair-breaking by disorder does not play a dominant role and that the mechanism of superconductivity in overdoped cuprate superconductors is qualitatively different from conventional mean-field theory.**

The essence of high-temperature superconductivity in the cuprates revolves around doping a Mott insulator. Superconductivity emerges when hole-doping is greater than 5% per lattice site; $T_c$ initially increases through the underdoped (UD) region of the phase diagram, before it decreases again in the overdoped (OD) region[1]. Superconductivity breaks down completely at roughly 27% doping. For the strongly overdoped region (SOD), it is often assumed that screening sufficiently reduces electron-electron correlations for a Fermi liquid to appear[5–7]. The superconducting state is then of the Bardeen-Cooper-Schrieffer (BCS) type, and the suppression of superconductivity is a consequence of a diminishing pairing interaction. Evidence for such conventional behavior in the OD regime comes from photoemission experiments, which suggest the existence of a full Fermi surface with superconductivity, as indicated by an energy gap that opens up in a BCS fashion below $T_c$[8,9]. As a caveat, very recent magneto-transport experiments indicate that even at high doping the normal state has strange metal features[10].

The first surprise in this regard was the discovery that the superfluid density decreases linearly to zero with doping beyond optimal doping[2,11,12], contrary to the BCS expectation that it should be of the order of the total carrier density and hence proportional to the doping level[1,2]. Additionally, optical conductivity measurements revealed a large density of metallic carriers[11], suggesting a filling of the superconducting gap due to pair breaking. One possible explanation for these observations involves disorder at length scales comparable to the small coherence length that is typical for the cuprates[3,4,13]. According to Bogoliubov-de-Gennes (BdG) theory (i.e. BCS in spatially heterogeneous systems), disorder at these length scales leads to emergent granular superconductivity[4,14–18], i.e. puddles of well-developed superconductivity with a size set by the coherence length, separated by a metallic matrix where the gap is suppressed. The resulting weak-link superconductor will show a low superfluid density.

We investigate these issues using scanning tunneling spectroscopy which yields the spatial distribution of the electron density of states with atomic-scale precision. Our measurements show that in $(Pb,Bi)_2Sr_2CuO_{6+\delta}$ (Bi2201) which has one $CuO_2$ layer per unit cell, such a "puddled" superconductor does indeed develop at high doping (Figs. 1,2). The typical spatial extent of the puddles is a few nanometers, of the order of the small coherence length in this system (Fig. 3). Our measurements additionally reveal that the superconducting gap persists beyond the dome, and that instead the heterogeneity is driven by gap filling (Fig. 4). This strongly suggests that the breakdown of superconductivity is not a result of a vanishing pairing interaction. A comparison with BdG simulations suggests that this filling is likely due to the decay of the Cooper pairs in surrounding metallic areas, which in turn explains the observation of a large density of metallic carriers. Unexpectedly, we also find a striking violation of a basic BdG rule. Within BdG theory, pair breaking goes hand-in-hand with gap closing, because depletion of the number of Cooper pairs in a superconductor leads to a diminishment of the gap magnitude Δ as well: $\Delta = V \sum_k \langle c_{k\uparrow}^+ c_{k\downarrow}^+ \rangle$, where $V$ is the attractive interaction and the $c^+$'s are electron field operators. Instead, our data show that the puddles characterized by the largest gap magnitudes exhibit also the largest gap filling (Fig. 4c), and that the average gap magnitudes are barely depending on doping (Fig. 4a). We therefore conclude that the physics governing the superconducting transition is of a different, non-mean-field kind.

To arrive at these findings, we study a series of Bi2201 samples with eight different doping levels, from underdoped to beyond the superconductor-metal transition, with an emphasis on the strongly overdoped regime. We chose Bi2201 because it has only one $CuO_2$ plane per unit cell, and has a rather large residual resistivity[19], suggesting that disorder is exceptionally important. On each sample, we measure the atomic-scale-resolved differential conductance $g(E,r)$ as a function of bias energy $E$ and location $r$, which is proportional to the Bogoliubov quasiparticle density of states.

We first consider the spatially averaged $g(E)$ spectra obtained at 4.2 K (Fig. 1a). Consistent with earlier reports[20–23], crossing into the overdoped regime, the spectra acquire an increasingly large non-zero Bogoliubov quasiparticle density of states at the Fermi level. This is remarkable as this quantity should go to zero for a standard d-wave BCS superconductor, but it is consistent with results from optical conductivity measurements[11]. It remains to be seen whether ARPES, if performed in SOD regions with such small gaps, would observe a similar phenomenology both in Bi2201 and Bi2212. We investigate this phenomenology using individual spectra, as in a heterogeneous situation like this, the average spectra do not represent the phenomenology adequately (Figs. 1b-d).

Next, we use a phenomenological model to fit all spectra over the whole doping range to extract the superconducting gap and gap filling of each individual spectrum. We calculate the spectral weight on each point $k = (k_x, k_y)$ on the Fermi surface using a Dynes formula with superconducting gap $\Delta_k = \Delta(\cos(k_x) - \cos(k_y))/2$, where $\Delta$ is the maximal gap, and then average over the Fermi surface[24]. We use the Dynes formula [25–27] as a mere phenomenological description constructed to reveal characteristic scales for the gap size and the gap filling, and discuss interpretational concerns after the presentation of the data. Our model yields the following function for the modeled differential conductance:

$$g(E) = P(E) * \langle \text{Dynes}(E, \Delta_k, \Gamma) \rangle_{FS} = P(E) * \left\langle \text{Re}\left(\frac{E + i\Gamma}{\sqrt{(E + i\Gamma)^2 - \Delta_k^2}}\right)\right\rangle_{FS} \quad (1),$$

where $\langle \rangle_{FS}$ indicates the average over the Fermi surface, $P(E)$ is a third-degree polynomial function to account for background variation, and Dynes($E$, $\Delta_k$, $\Gamma$) is the Dynes function with the pair-breaking parameter $\Gamma$. For this study, we concentrate on the superconducting gap and thus restrict ourselves to a ±15 meV range (in the underdoped and optimally doped range, which are not the focus of this paper, a pseudogap exists at a larger energy scale, as indicated by the red arrows in Fig. 1a). Lastly, we convolute $g(E)$ with a Gaussian function to account for spectrum broadening due to a finite temperature and the lock-in modulation.

We define the filling parameter $F$ as the ratio $g(E = 0, T \to 0)/P(E = 0)$, which can be expressed in terms of our fitting parameters as:

$$F = \langle 1/\sqrt{1 + (\Delta_k/\Gamma)^2} \rangle_{FS} \quad (2).$$

Figs. 2a,b illustrate how the model differentiates between gap closure, controlled by $\Delta$, and gap filling, controlled by $\Gamma$ or $F$. Fig. 2c shows some typical spectra and fits from various locations. It is clear that, when compared to the scenarios presented in Figs. 2a,b, the measured spectra look more similar to the filling scenario as opposed to the closing scenario. We then fit roughly $10^5$ spectra from 8 different doping levels with this model, and display the extracted gap size and gap filling in Fig. 3. We note that for the strongly overdoped samples, high signal-to-noise is key for successful fits; the traces shown in Figs. 1b-d and Fig. 2c are raw spectra without any averaging. A further challenge is that at higher doping, a significant fraction of spectra exhibit completely gapless regions. We identify such spectra after fitting and exclude them from subsequent analysis. In the supplementary information, we provide details (see Supplementary SIIB and figure S4) and demonstrate that our key

results are independent of these choices (Supplementary SIIA-C). We also provide a modified version of the model with an alternative definition of the gap filling, and show that our results are independent of the precise definition of gap filling (Supplementary SIID).

We start our discussion with the spatial maps of the gap size $\Delta(r)$ as a function of doping (Figs. 3b-d). Strikingly, while more spectra are fully filled at higher doping, the average gap size remains roughly constant on the strongly overdoped side (Fig. 3a). Initially, the gap size increases when moving from underdoped to optimally doped. Beyond optimal doping, the gap size barely decreases anymore when going through the OD and SOD side, and instead remains roughly constant – even beyond the superconductor-to-metal transition. In particular, throughout the SOD region, we observe an almost constant average gap amplitude even though $T_c$ is rapidly decreasing. Our study thus excludes a homogenously diminishing pairing interaction as the cause of the superconductor-to-metal transition.

Given a constant gap, what drives the changes in spectra on the overdoped side? Our analysis indicates that it is the gap filling. We extract the gap filling, $F$, using eq (2), for each measured sample, and present the distribution of the gap fillings in Figs. 3f-h, and their histograms in Fig. 3e. Remarkably, the mean gap filling changes considerably over the doping range. In the OP region, the spectra have a filling close to zero, i.e. they are fully gapped. Crossing into the OD regime, a subset of spectra starts to develop a finite gap filling. This subset grows with further doping, with all spectra having a finite filling in the SOD regime. The values of $F$ shift markedly in this doping regime from nearly fully gapped ($F = 0$) near optimal doping to almost fully filled ($F = 1$) towards the SOD regime and extending into the metallic regime. The trends in gap closing and gap filling are summarized in Fig. 4a,b: as the doping is increased into the overdoped regime, the gap size remains roughly constant; in contrast, the gap filling increases rapidly. Thus, a first key result of this paper is that it is not a decaying gap width $\Delta$, but an increasing gap filling $F$ that is responsible for the diminishing superconductivity and eventually drives the superconductor-to-metal transition.

Notably, the gap filling is highly heterogenous, as can be seen from the width of the distributions in Figs. 3e and in the spatial maps in Fig. 3f-h. We observe areas both with and without a gap, each existing at a length scale consistent with the coherence length (~ 1.5 nm). This observation indicates that the breakdown of superconductivity in the overdoped regime of the single-layer bismuth cuprate is likely caused by an emergent strongly inhomogeneous superconductivity, leading to an effective weak-link physics that explains the diminishing superfluid density. Hence, at first glance, our data suggest that theoretical models involving disorder-driven breakdown of superconductivity in the BdG framework[3,4,13–15] are a good description of the physics of strongly overdoped Bi2201, with the additional information that it is the gap filling that drives the formation of the superconducting puddles.

Next, we focus on the origin of the gap filling. According to BdG theory, the excitations that fill the gap are quasiparticles of the Fermi-liquid normal state that are released by breaking up Cooper pairs. Well-known causes for pair breaking are potential disorder[3,4,13] (for a d-wave superconductor) and thermal phase fluctuations[28–31]. However, if potential disorder were the only culprit, the areas where the pair breaking is smallest (where superconductivity survives best) should have the largest gaps, which is not what we observe. We demonstrate this in figure 4c, where we show the local relationship between the gap size $\Delta$ and the pair breaking $\Gamma$, and find a clear positive correlation between the two. Further, we can exclude thermal phase fluctuations based on our temperature dependent measurements, up to 20 K for the OD9K sample. Thermal phase fluctuations should lead to a strongly temperature-dependent filling, in contrast to our observations (see supplementary SIII).

We therefore consider an alternative candidate for pair breaking: the decay of Cooper pairs into smaller gap or metallic regions, as previously suggested[32–34]. This can be seen as akin to an inverse proximity effect[35]. We are not aware of self-consistent simulations for this scenario in the literature, but they are possible with state-of-the-art numerical methods. We start with a large real-space supercell implementing a realistic tight-binding band structure. We then introduce the superconducting puddles by switching on a local pairing interaction characterized by a linear dimension $L$ that is approaching the (bulk) coherence length. The BdG equations are then solved self-consistently (see supplementary SIV for further details) and typical outcomes are shown in Fig. 1e. The simulated spectra are surprisingly similar to the experimental ones, and one might wonder whether this gap-filling dominated physics is connected to certain disordered superconductors[36,37] and interface superconductors[38] with a local density of states phenomenology that is not dissimilar from what we observe here.

However, there is one aspect of our data that is markedly inconsistent with the BdG description of granular superconductors. Our data shows that the largest gaps also exhibit the strongest gap filling (Fig 4c), while within BdG, gap filling should always go hand-in-hand with a decrease of the gap magnitude. Our self-consistent simulations confirm that this is indeed also valid for the heterogeneous case: upon application of our fitting model to the calculated spectra shown in Fig. 1e, we find that the regions with the largest gaps show the least amount of pair breaking as shown in Fig 4d. We note that self-consistency in the calculations is necessary here; fixing the gap magnitude artificially would obscure any effect from pair breaking on the gap magnitude. The comparison between Fig 4c and 4d shows a striking inconsistency between the experiment and BdG expectation.

In summary, our real-space imaging reveals a strongly heterogeneous superconductivity consisting of superconducting puddles with a size set by the coherence length immersed in a metallic matrix. This explains the diminishing superfluid density[2] and the origin of the large fraction of metallic carriers[11]: it stems from the filling of the gap. Our data further demonstrate that superconductivity does not, as it is often assumed, become conventional in the SOD regime. The breakdown of superconductivity is not the consequence of a vanishing pairing interaction and does not follow the BdG description. Furthermore, what fills the gap is entirely different from simple quasiparticles that count the number of broken Cooper pairs. Instead, the gap filling might be related to the strange normal state [10], e.g. collective excitations of an unknown kind rooted in the "strange metal" physics, which at present cannot be calculated, or electrons from a different sector that does not exhibit a spectral gap, but not as a consequence of pair-breaking. Last but not least, this unconventional physics may not be limited to the low-temperature, overdoped regime. STM studies at optimal doping of Bi2212 showed a rather similar puddling effect upon approaching the superconducting transition temperature[34,39]. This may imply that the physics of the thermal transition – the "high $T_c$" problem itself – is governed by unknown physics. It would be interesting to revisit this high-temperature regime to make this more precise.

*Author Contributions*
W.O.T, T.B., J-F.G., K.M.B, D.C., & Y.Y performed STM experiments. S.S. & M.S.G performed ARPES measurements. W.O.T, I.B., & A.V. did data analysis. J.Z & M.A.S. performed the theoretical simulations. W.O.T., T.B. J-F.G., J.Z., & M.P.A. wrote the manuscript. Y.H., T.K., & E.vH made and characterized the samples. M.P.A. supervised the project. All authors commented on and agreed on the manuscript.

*Acknowledgments*
We thank Eric Hudson, Doohee Cho, Remko Fermin, Evert Stolte, Lewis Bawden, Koenraad Schalm, Maarten Berben, Nigel Hussey, Gijsbert Verdoes, Kees van Oosten, & Freek Groenewoud for their help and insightful discussions. This work was supported by the European Research Council (ERC StG SpinMelt). M.A.S. acknowledges funding support from NSF-DMR-1849751 and the IARPA SuperTools program.

*Data Availability*

The data used to generate the figures in the main text and supplementary information will be available on Zenodo. All other data generated during this project is available upon reasonable request to the authors.

*Code Availability*

The code used for this project is available upon reasonable request to the authors.

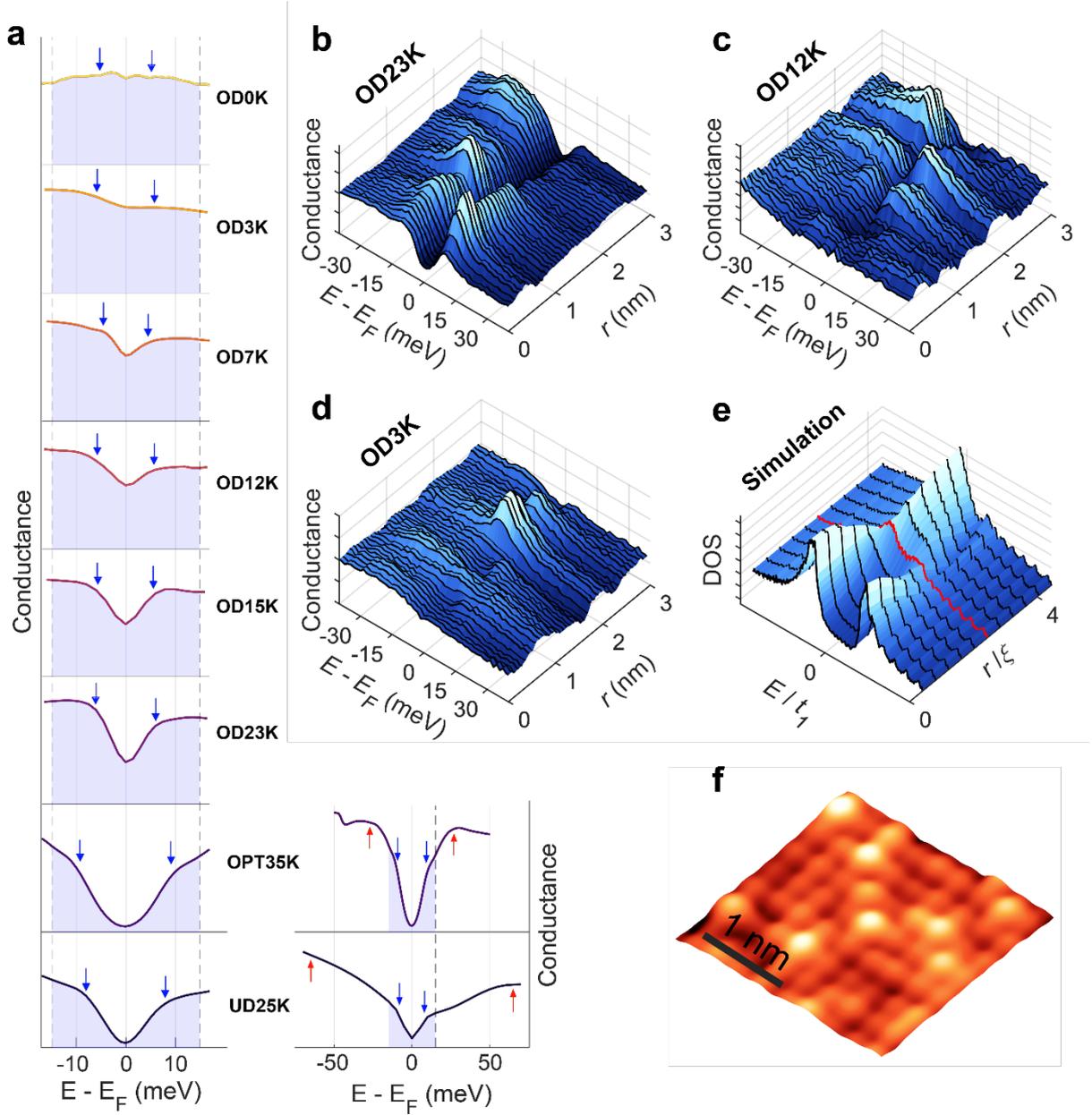

**Figure 1: Heterogeneous gap filling in Bi2201. a** The average spectra of eight different doping levels, labeled by their $T_c$. The shaded areas indicate the energy range used in the fitting procedure. The blue arrows show the average extracted gap magnitude. For the UD25K and OPT35K samples, the red arrows indicate the pseudogap as determined by He et al.[22]. **b-d** Spectra along 3 nm linecut for the OD23K, OD12K, and OD3K samples respectively. These raw, unprocessed spectra indicate the high degree of electronic inhomogeneity in these samples. **e** A linecut of spectra from a self-consistent BdG simulation from the center of a superconducting puddle ($r = 0$) to the metallic environment which shows the van Hove singularity modeled to be close to the Fermi level (see main text and supplementary SIV for details). The boundary of the puddle is indicated by the red spectrum. The energy unit is relative to the hopping parameter $t_1$, and the length unit is relative to the coherence length ξ (see supplementary SIV). **f** Typical topography measured on the OD12K sample on the same length scale as figs 1b-d.

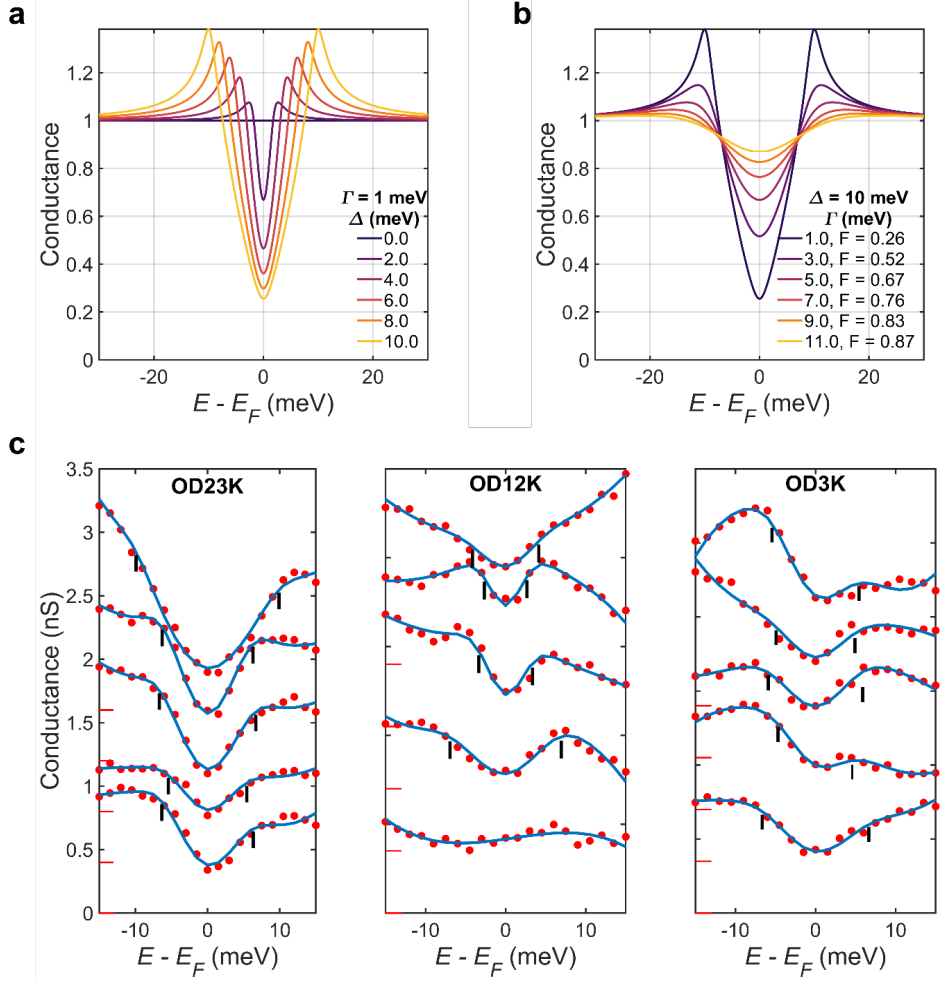

**Figure 2: Gap filling versus gap closure.** Difference between gap closure (a) and gap filling (b) by presenting a sweep of the gap magnitude parameter $\Delta$ for constant $\Gamma$, and a sweep of the scattering rate parameter $\Gamma$ for constant $\Delta$, respectively. **c** Example fits from our model applied to our raw data for the OD23K, OD12K, and OD3K data. The zeros of the spectra are offset for better visibility, as indicated by the red marks. The black marks indicate the gap width as determined by the model.

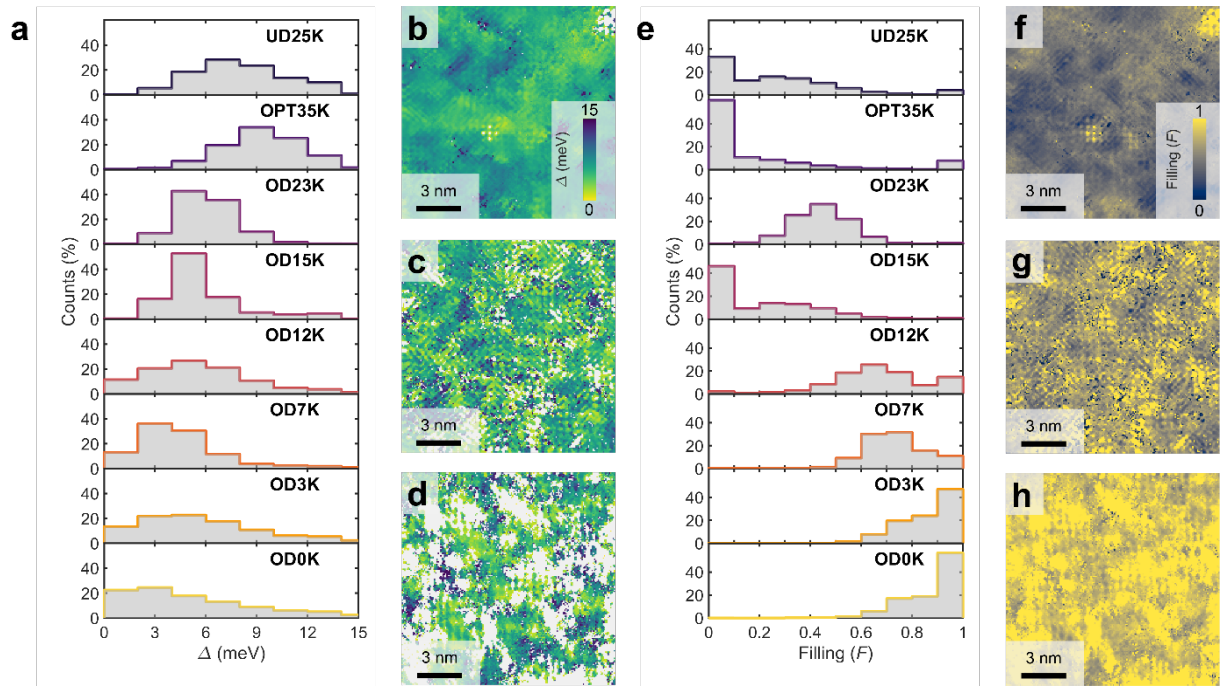

**Figure 3: Doping dependence of the spatially resolved gap filling and gap magnitude. a** Gap magnitude histogram for each doping concentration. **b-d** The spatial distributions of the gap magnitude for the OD23K, OD12K, and OD3K samples. The spectra that are omitted from the histograms (see main text and supplementary SII) are indicated by the white areas in figure b-d. **e** Gap filling histogram for each doping level. **f-h** Spatial distribution of the gap filling for the OD23K, OD12K, and OD3K samples respectively.

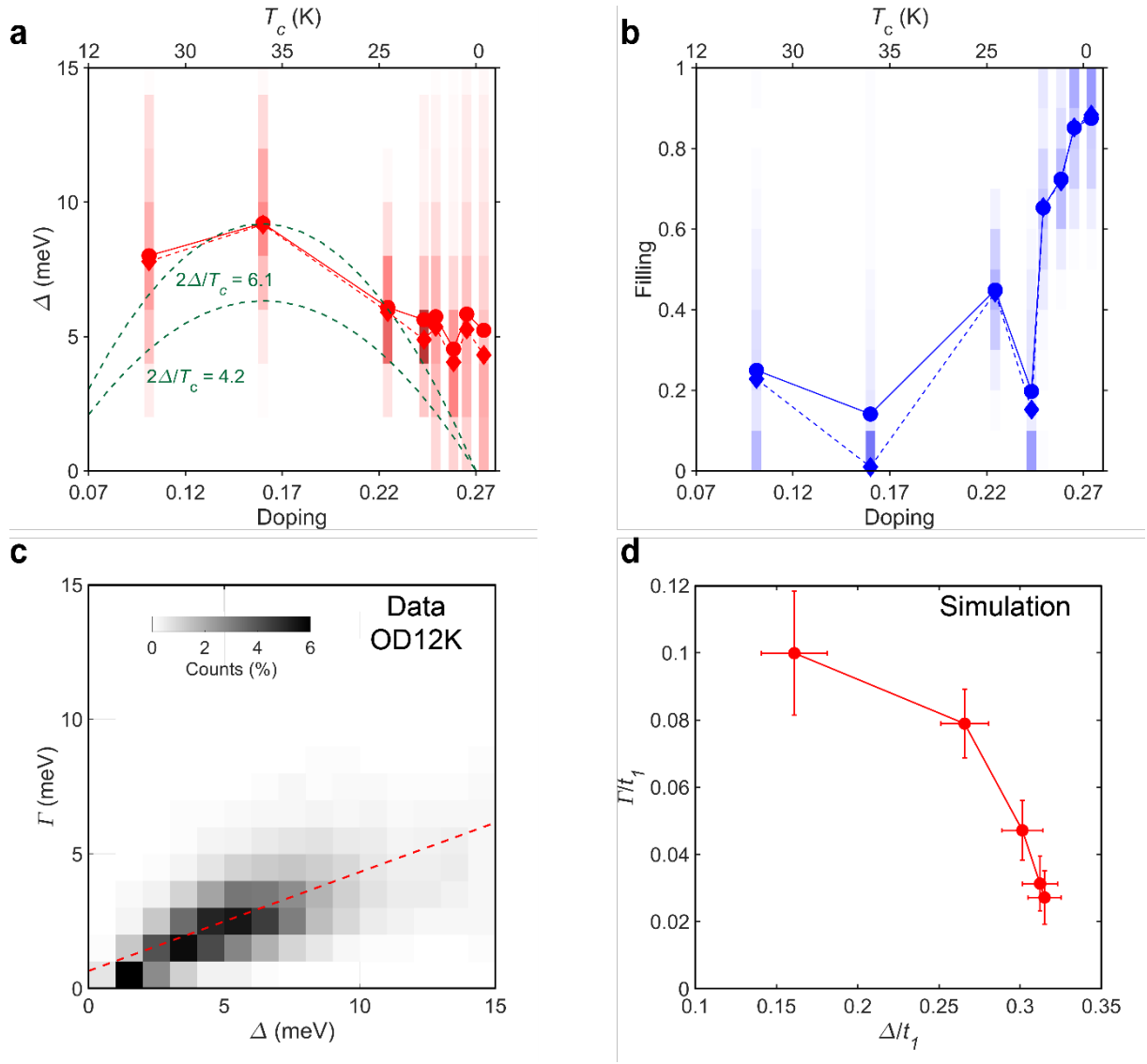

**Figure 4: Gap filling driven breakdown of superconductivity and the contradiction to BdG. a,b** The mean (circles) and median (diamonds) of the gap magnitude and the gap filling, respectively. The shaded areas represent the local variations in the gap magnitudes and fillings by depicting the histograms. The green dashed lines in **a** show the expectation of a gap size proportional to $T_c$, with a proportionality constant either chosen to match the OPT35K data point, or determined by weakly coupled d-wave BCS theory. **c** 2D histogram of the measured local relation between gap magnitude $\Delta$ and pair-breaking parameter $\Gamma$ for the OD12K sample. The positive correlation between the two is indicated by the red line. **d** The relation between the gap magnitude and pair-breaking parameter extracted from self-consistent BdG simulations (Fig. 1e) using the same fitting model. In contrast to figure c) we find a clear negative correlation. The error bars indicate the uncertainty in the obtained values due to the fitting process.

# Supplementary Information

## I. Experimental Methods

We performed a systematic study on a series of $(Pb,Bi)_2Sr_2CuO_{6+\delta}$ samples with 8 distinct doping levels, covering the range from underdoped (UD25K) to strongly overdoped (OD0K) side. Single crystal samples were grown by the conventional floating-zone technique[1,2]. The UD25K, OP35K, and OD15K samples contain La doping, i.e. $(Pb,Bi)_2(La,Sr)_2CuO_{6+\delta}$, while the rest of the samples are without La doping. The doping levels, transition temperatures $T_c$, and measurement temperatures are listed in Table S1. The doping levels of the superconducting samples are determined using the Presland formula, while the doping level of the OD0K sample is extracted from the rigid band shift measured by ARPES (see supplementary section V). All samples were cleaved *in situ* in a cryogenic environment and inserted immediately into the STM. The data were acquired using different home-built cryogenic STMs among three groups.

| Name | Doping | $T_c$ (K) | Measurement $T$ (K) | Data acquired by | Samples fabricated by | $V_{setup}$ (mV)/ $I_{setup}$ (pA)/ Lock-in Amplitude (mV) | Effective Energy Resolution (FWHM, meV) |
|---|---|---|---|---|---|---|---|
| UD25K | 0.101 | 25 | 5.7 | Hudson group | Kondo group | -100/400/1 | 2.7 |
| OP35K | 0.160 | 35 | 5.7 | Hudson group | Kondo group | -100/400/1 | 2.7 |
| OD23K | 0.224 | 23 | 4.2 | Allan group | UvA group | -150/150/1.5 | 2.58 |
| OD15K | 0.243 | 15 | 6 | Hoffman group | Kondo group | -100/100/2 | 3.5 |
| OD12K | 0.249 | 12 | 4.2 | Allan group | UvA group | -200/170/1.5 | 2.58 |
| OD9K | 0.255 | 9 | 4.2 – 20 | Allan group | UvA group | -150/200/1.5 | 2.58 – 6.42 |
| OD7K | 0.258 | 7 | 2.2 | Allan group | UvA group | -20/600/1.5 | 2.50 |
| OD3K | 0.265 | 3 | 4.2 | Allan group | UvA group | -200/170/1.5 | 2.58 |
| OD0K | 0.274 |  | 4.2 | Allan group | UvA group | -25/200/1.5 | 2.58 |

**Table 1:** Samples and their measurement conditions in this study.

## II. The phenomenological model to fit spectra

### A. The *d*-wave gap model

The *d*-wave gap is here modeled as a mean of multiple *s*-wave gaps, one for each point along the Fermi Surface. For each *s*-wave gap, the gap size $\Delta_\mathbf{k}$ is given by $\Delta_\mathbf{k} = \Delta(\cos k_x - \cos k_y)/2$. The points $k_x, k_y$ are found using the Fermi Surface of the tight-binding model for the OD15K sample[1] although the final shape of the spectrum varies a lot with the exact points in k-space used. Each s-wave gap is generated by using the Dynes formula

$$\text{Dynes}(E, \Delta_{\mathbf{k}}, \Gamma) = \text{Re}\left(\frac{E+i\Gamma}{\sqrt{(E+i\Gamma)^2 - \Delta_{\mathbf{k}}^2}}\right),$$

where the same $\Gamma$ is used for all *s*-wave gaps. The resulting *d*-wave gap is the mean of all *s*-wave gaps. To account for the normal-state density of states (DOS), the *d*-wave gap function is multiplied with a polynomial function, typically of the 3rd order. The resulting spectrum is then convoluted with a Gaussian function with a full width at half maximum (FWHM) given in the table above, in order to emulate the effect of finite temperature and the lock-in modulation have on the shape of the measured spectrum.

The points in momentum space are calculated only once before the fitting process to reduce computation time. The fitting parameters characterizing the gap are only $\Delta$ and $\Gamma$. To calculate the filling for the *d*-wave gap we calculate the mean of the filling for each individual *s*-wave gap using $F = [1 + (\Delta/\Gamma)^2]^{-1/2}$.

## B. Statistical analysis with and without the excluded spectra

In the main text we "white out" certain spectra (white areas in Figs. 3b-d), and exclude them for the statistical analysis when either of the two conditions is met in the fit results: 1) $\Gamma$ > 20 meV; 2) $\Delta$ > 15 meV. Our interpretation is that these spectra are fully filled, for the following reason: the spectra that meet the first criterion have so much broadening that there is no well-defined gap. Similarly, for the spectra that meet the second criterion, the large "gap" is a reflection of background modulations. Such spectra are thus counted as fully filled. Still, we show here that including these spectra in the analysis does not alter our main conclusions.

Figure S1 shows, from left to right, the spatial distributions of: $\Delta$, including "whited-out" (1st column) and excluding "whited-out" (2nd column) spectra, $\Gamma$, with (3rd column) and without (4th column) "whited-out" spectra, and $F$, the filling (5th column) for all samples. The images are ordered top to bottom, from lowest to highest doping, respectively. Following the argumentation in the preceding paragraph, the filling for "whited-out" spectra is set to 1, when they are included in the statistics. Figure S2 shows the histograms for $\Delta$ including "whited-out" spectra, and the histograms of $\Gamma$, both with and without "whited-out" spectra. The remaining histograms of $\Delta$ excluding the "whited-out" spectra and of the filling can be found in the main text. The histograms in Fig. S2 are summarized in Fig. S3 in a similar fashion to Fig. 4 of the main text.

From the spatial distributions and histograms of $\Delta$ and $\Gamma$ in Figs S1-S3, we conclude that even when the "whited-out" spectra are considered: 1) the gap size still deviates from the $\Delta \propto T_C$ behavior in the OD regime; 2) gapped spectra can still be found in the non-SC sample in significant quantities.

The spatial averages of the "whited-out" spectra and of the rest spectra are shown in Fig. S4 for each sample. We find that after whiting out all samples, even the non-SC sample, show a gap in their average spectrum. We note that the spatially-averaged "whited-out" spectra in the OD regime are fully filled, i.e. they no longer have a gap, and often show a peak near Fermi level. This further justifies our choice to assign all these spectra a filling of 1. For the UD25K and OPT35K samples, the "whited-out" spectra are made up of spectra for which the fit has failed due to limited signal-to-noise. Even though these spectra appear as gapped, we attribute this to the presence of a pseudogap. We find that the assignment of $F = 1$ to these spectra does not alter the main conclusion either, given the relatively small portions of "whited-out" spectra in these samples (see Fig. S4). The increase of the area of "whited out spectra" in the OD as shown in Fig. S4 reaffirms the increased gap filling in these samples.

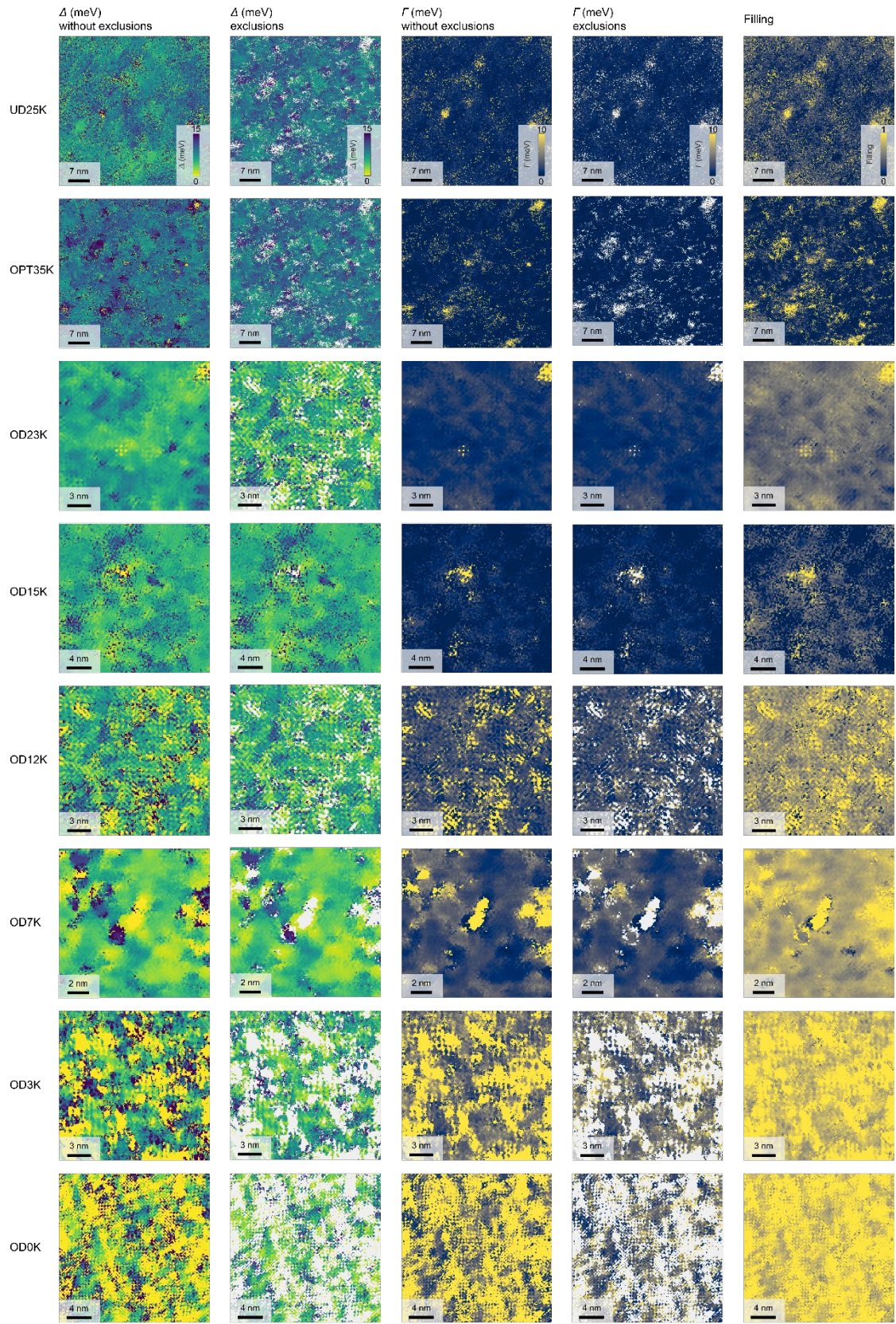

**Figure S1:** Spatial variations of the gap size $\Delta$ with and without exclusions, parameter $\Gamma$ with and without exclusions, and the calculated filling for all samples. See text for details.

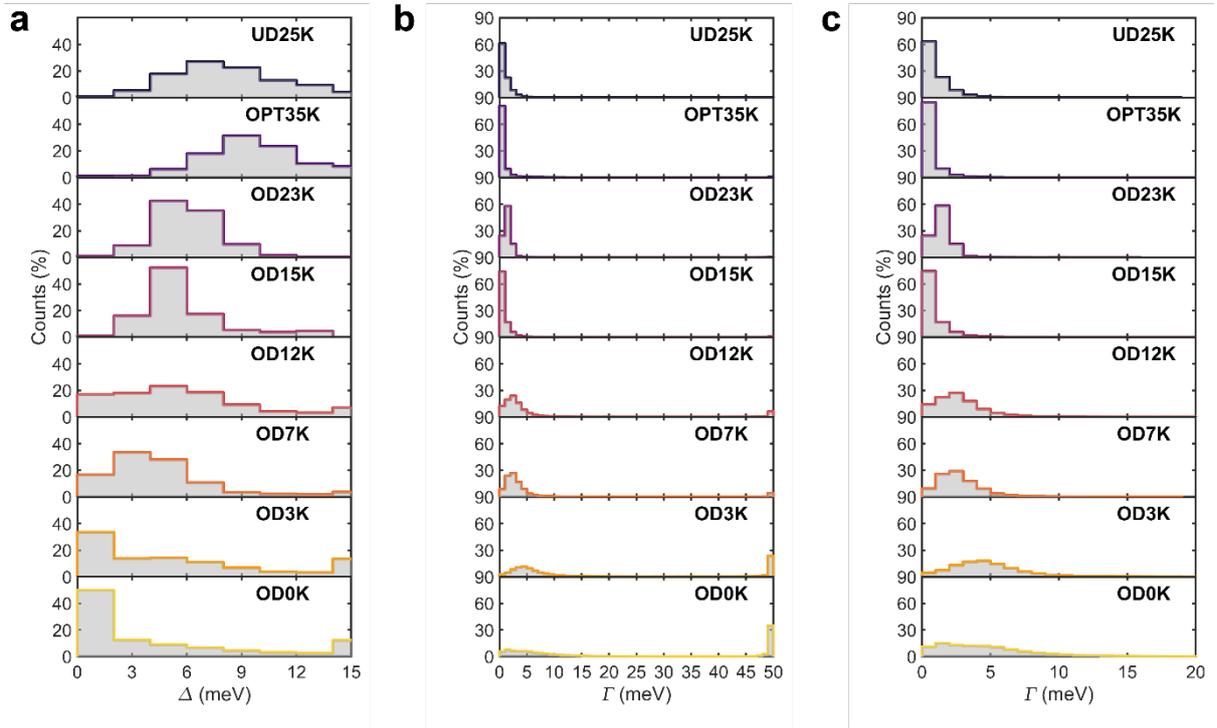

**Figure S2:** Histograms of the gap size Δ *without* any spectra that are whited out (a). Histograms for Γ including (b) and excluding whited-out spectra (c). See text for details.

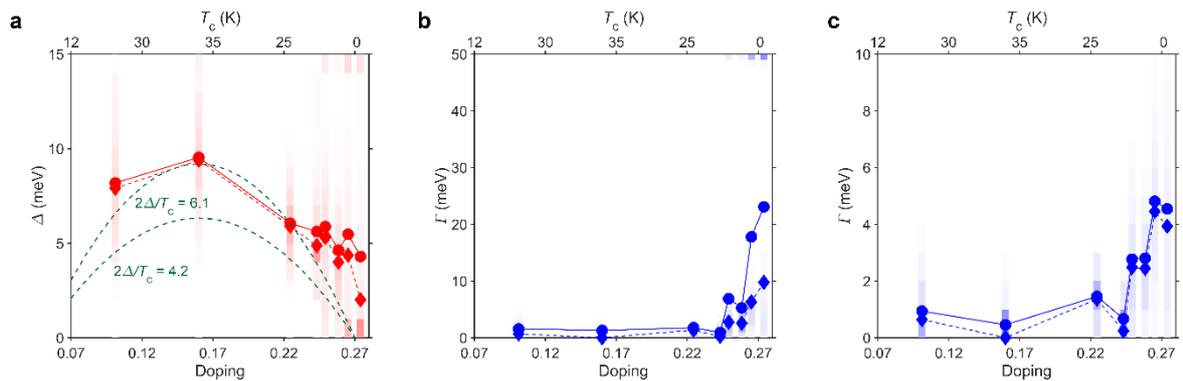

**Figure S3:** The results for Δ without excluding any spectra, and for Γ including and excluding "whited-out" spectra are summarized (a, b, c respectively). The circles indicate the mean Δ, Γ for each sample, with the diamonds indicating the medians. The shaded areas in the background represent the spread in values these parameters have. The green dashed lines in the left figure indicate the behavior expected for Δ proportional to $T_c$. The value of $2\Delta/T_c$ corresponds to the dirty d-wave BCS limit, while $2\Delta/T_c$ is chosen such that it matches the OPT35K data point.

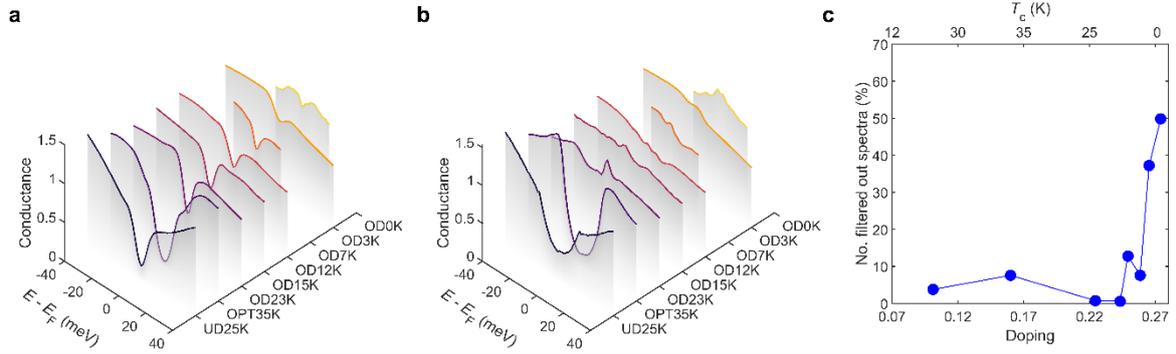

**Figure S4: a)** The spatially-averaged spectra after the "whited-out" spectra have been removed. **b)** Spatial averages of the "whited-out" spectra for each doping level. **c)** The proportions of "whited-out" spectra that make up each sample as a function of doping.

### C. Energy range for fitting and approximations for the normal density of states

The spectra in the UD and OPT samples show clear pseudogap (PG) features, with PG sizes ranging from 20 meV to over 60 meV (see main text, figure 1a). Furthermore, in the OD regime, the normal-state DOS shows a peak near the Fermi level. These additional features next to the superconducting gap complicate the accurate fitting of the superconducting gap. We circumvent this complication by limiting our analysis to a small window $E_{win}$ around the Fermi level. In this reduced energy window, the additional features are only partly visible, and can be sufficiently approximated by a polynomial DOS. The choices of $E_{win}$ and the order of the polynomial background are arbitrary but necessary choices made before the fitting procedure. Here we show the influence the particular choices have on the superconducting gap size and filling in the OD samples.

In the main text, we use $E_{win}$ = ±15 meV, which is a choice made before the fitting procedure. Altering this choice does not affect the main conclusions of our analysis, as shown in figure S5. We repeat the analysis using different energy windows and find that the qualitative behavior does not change: $\Delta$ remains constant while $F$ sharply increases in the SOD regime. Further increasing $E_{win}$ beyond 20 meV, the highly inhomogeneous normal-state DOS becomes more significant, defeating the aim of focusing on the superconducting gap through an energy window. With an energy window smaller than 10 meV, we find that too little of a spectrum is left to characterize the superconducting gap accurately.

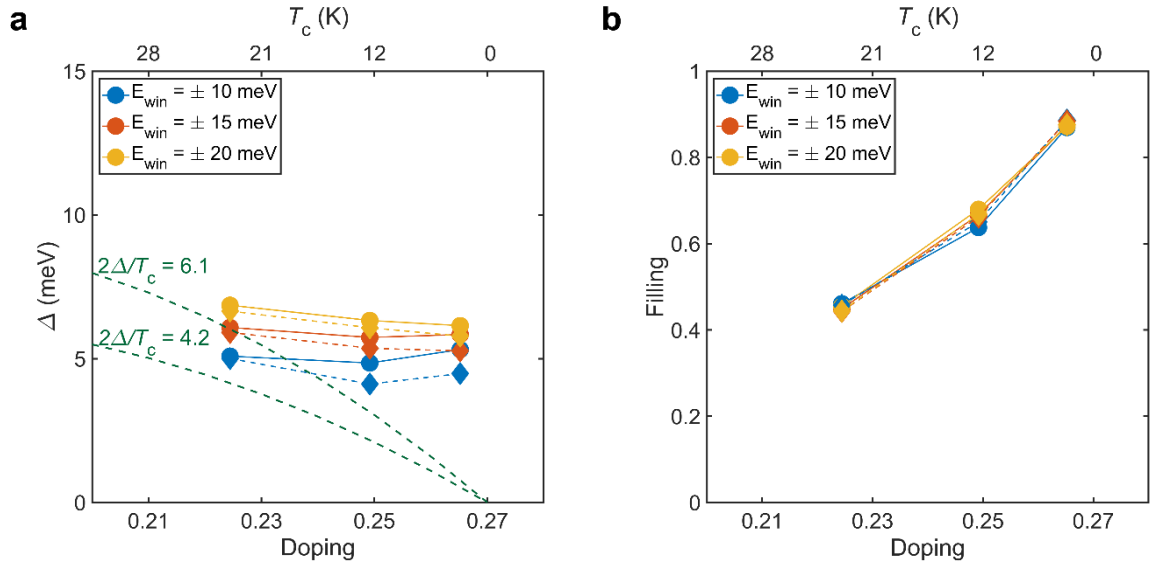

**Figure S5:** Dependence of our conclusion on the choice of the energy window. The dependencies on the fitting energy window for the average $\Delta$ and filling (a, b respectively) for the overdoped samples OD23K, OD12K, and OD3K. While the absolute values of the averages vary slightly with the cutoff energy, the overall behavior of a constant gap size and increasing filling factor is independent of the cutoff.

Another possible influential choice in the fitting procedure is the order of the background polynomial used to model the normal-state DOS. Figure S6 shows the mean gap size and filling for the OD23K, OD12K, and OD3K samples for different orders of polynomial ranging from 1$^{st}$ to 4$^{th}$ order. The overall behavior of nearly constant gap size and increasing filling is present for all polynomial orders. We opt to use a 3$^{rd}$ order polynomial in the main text as it offers the best balance between underfitting and overfitting.

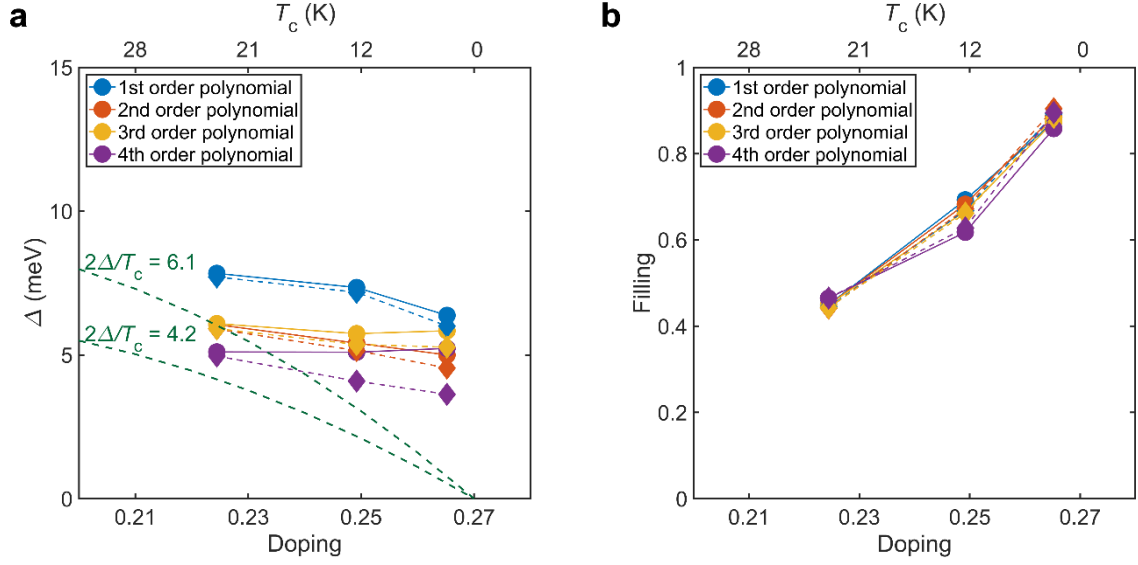

**Figure S6:** Dependence of our conclusion on the choice of the polynomial normal-state DOS. The average gap size and filling (a, b) for the overdoped samples OD23K, OD12K, and OD3K as the order of the polynomial normal-state DOS is varied. While the absolute values of the averages vary a bit among the various polynomials, the overall qualitative behavior of a barely varying gap size and the drastic increasing filling is present in all cases.

### D. An alternative model

Here we introduce an alternative approach to determining the gap filling, and show that the conclusions are the same using this model. We use a model which explicitly includes the filling *F* as a fitting parameter, in contrast to extracting F using fit parameters *Δ* and *Γ* in the main text:

$$g(E) = P(E) * [(1 - F) * \text{Dynes}(\Delta, \Gamma) + F],$$

where $P(E)$ and $\text{Dynes}(\Delta, \Gamma)$ are defined the same as those in the main text. The gap filling is now explicitly parametrized by the parameter *F*, with *F* = 0 corresponding to fully gapped and *F* = 1 to fully filled. To prevent overfitting and to limit built-in correlations between fit parameters we fix the value of *Γ*. Figure S7 shows the average gap size and gap filling from the fit results using this model, analogous to Fig. 4 in the main text. For this alternative model, we exclude spectra with a) *Δ* close to 0 (*Δ* < 1 meV), and b) *F* close to 1 (*F* > 0.95) from further analysis. In case a) the gap sizes become smaller than our thermally limited energy resolution, preventing an accurate determination of *Δ*. In case b) *F* becomes ill-defined as *F* can be absorbed into $P(E)$ when the gap is barely present. Fitting our data with this model, we find that the gap size remains constant in the OD regime, while the gap filling increases rapidly. Confirmation by an alternative model further strengthens the conclusions of the main text.

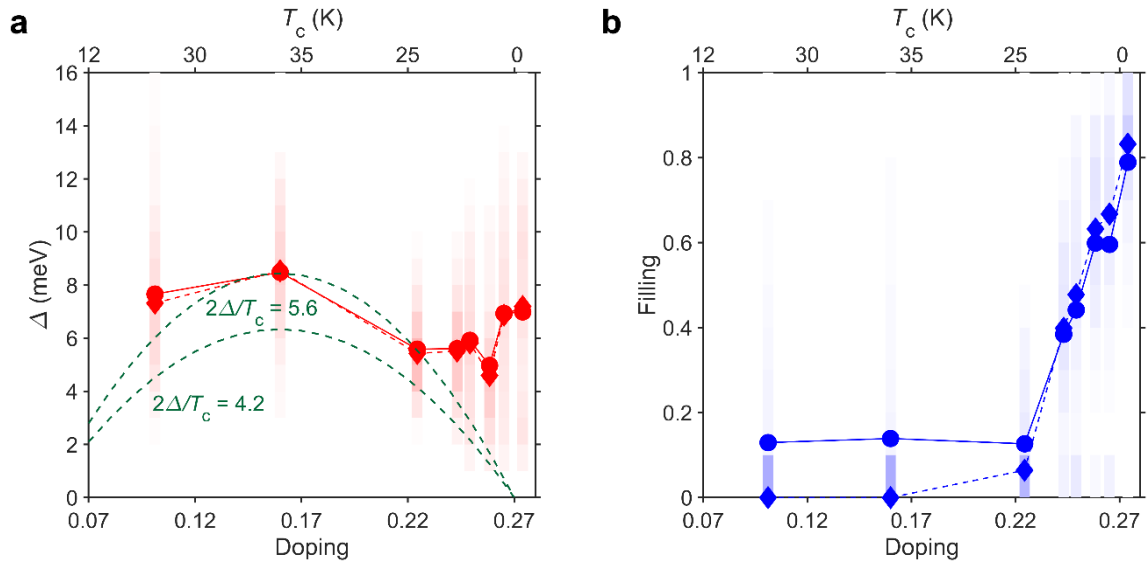

**Figure S7 Gap magnitude and filling versus doping using the alternative model.** The gap size (a) and the gap filling (b) as determined by the application of the alternative model described above. The shaded areas indicate the histograms of the parameters at each doping level. The average gap sizes and average filling are indicated by the circles. The median (diamonds) is shown to better reflect highly asymmetric distributions as is the case for the gap filling. The robustness of the qualitative trends against the use of different models reinforces the conclusions of the main text.

### III. Temperature dependence

In figure S9a, we show the temperature evolution of the average spectrum measured in the same field of view on the OD9K sample. Furthermore, we show the median values for the gap and filling parameters as a function of temperature in the same field of view in Figures S9b and S9c, respectively. We find that a gap is still present up to 20K for the OD9K sample, even when the temperature-limited and lock-in broadened energy resolution is taken into account. With increasing temperature, we see that the gap magnitude and gap filling remain fairly constant up to 20K.

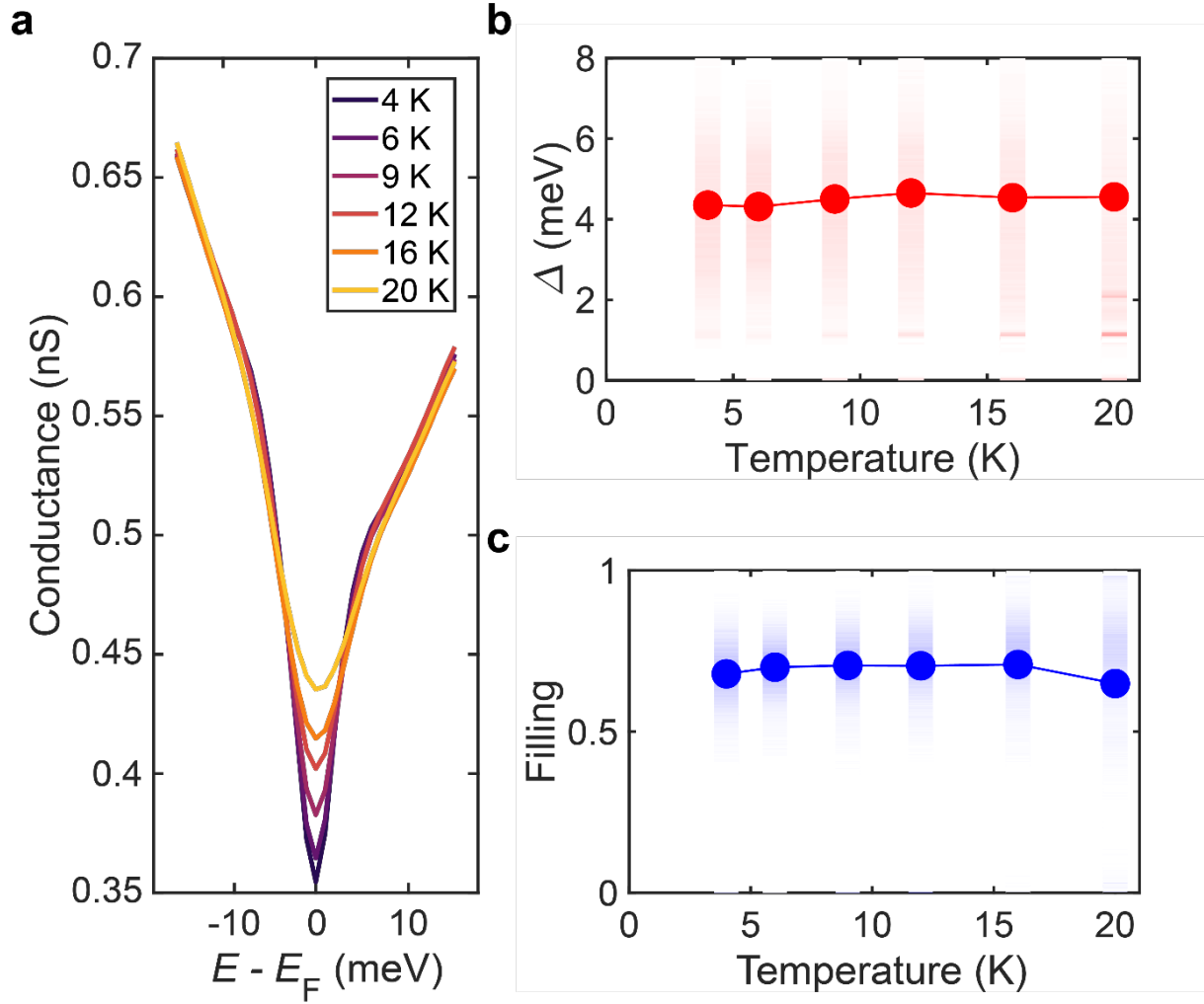

**Figure S8:** Temperature evolution of the gap width and gap filling. a) Average spectra measured in the same field of view on the OD9K sample (V = -150 mV, I = -200 pA). b) Median values for the gap magnitude in the same field of view of A, as a function of temperature. The shaded areas indicate the histograms of the gap at each temperature. c) Median values for the Filling parameter in the same field of view of A, as a function of temperature. The shaded areas indicate the histograms of the Filling at each temperature.

## IV. Intrinsic Metal-Induced Pair-Breaking Effects Within a Superconducting Puddle Embedded in a Metallic Matrix

In this section, we consider the case of a small *d*-wave superconducting puddle whose size is of the order of the superconducting coherence length $\xi_0$, embedded in a metallic matrix. This is a particularly relevant model for the strongly overdoped regime. Our treatment of this system is entirely mean-field; we employ large-scale numerical simulations of the Bogoliubov-de Gennes Hamiltonian to uncover interesting aspects of this system. To our knowledge, such calculations have not previously been performed in the literature; similar calculations (but for a superconducting puddle in a superconducting matrix) have been reported by Fang *et al.*[3] and Nunner *et al*[4].

We show here that one striking effect of the surrounding metallic matrix is to significantly *weaken d-*wave superconductivity, such that the resulting *d*-wave order parameter within the puddle is less than that of a bulk homogeneous system with the same pairing interaction. Decreasing the size of the puddle has the effect of decreasing the average *d*-wave order parameter within the puddle. We additionally find a concurrent increase in the average local density of states (LDOS) at zero energy within the puddle when the puddle size is decreased. The behavior of the puddle as its size is decreased is vastly reminiscent of the effect of increasing disorder on bulk *d*-wave superconductivity, and originates entirely from the mixing of the superconducting states within the puddle with the metallic states of the surrounding matrix. Thus, the surrounding metal induces pair-breaking within the puddle, very similar to the effect of disorder[5–7]. Importantly, our calculations confirm that the negative correlation between gap size and filling expected in the mean-field theory also holds for the heterogeneous case (see Fig. 4d in the main text).

In our simulations, we assume that the *d*-wave superconducting puddles are square patches of size *l* × *l*. We self-consistently calculate the order parameter from $\Delta_{rr'} = V_{rr'} \langle c_{r\uparrow} c_{r'\downarrow} \rangle$. The pairing interaction $V_{rr'}$ is nonzero only for nearest-neighbor bonds attached to sites within the *l* × *l* patch, and otherwise vanishes. To compute $\Delta_{rr'}$ and the LDOS $\rho(r, \omega)$, we employ an exact real-space Green's function method particularly suited for very large inhomogeneous systems[8,9]. For the calculations reported here, the system size is 100 × 200, which is larger compared to what more traditional exact-diagonalization methods can access. We iterate the calculation until the order parameter is converged, and we assume that we are at *T* = 0. We take the normal-state dispersion (up to next-nearest-neighbor hopping) to be given by the following parameters: $t_1$ = 1, $t_2$ = -0.33, and $\mu$ = -1.22 (from this point on we express all energies in units where $t_1$ = 1). The spatially resolved site-centered *d*-wave order parameter plotted throughout this section is obtained by adding the order parameter on all four bonds connected to a single site but assuming a sign difference between the order parameter on bonds along the *x*-direction and that on bonds along the *y*-direction.

We are interested in determining whether *d*-wave superconductivity in puddles behaves differently compared to the bulk case due to the abundance of low-energy states in the nearby metal, and we will tune the size of the puddles (from 5 × 5 to 17 × 17) in particular to isolate the effect of the nearby metal. One expects that the smaller the puddle, the stronger the effect of the metal, since a larger fraction of the puddle is in close proximity to the metal-superconductor boundary. We take the nearest-neighbor pairing interaction strength to be $V_0$ = 1.0 inside the superconducting puddle and $V_0$ = 0 outside it. To provide a baseline for comparisons, we perform the same calculation for a bulk *d*-wave superconductor as well, with the same nearest-neighbor pairing interaction $V_0$ = 1.0 present throughout the entire system. We will frequently express the puddle size in terms of $l/\xi_0$, where $\xi_0$ is the coherence length of the superconducting condensate within the puddle; this is to make clearer the effects of miniaturizing the superconducting puddle to dimensions of the order of $\xi_0$ itself. Note that we have suppressed here the explicit *l*-dependence of $\xi_0$, since as it turns out the magnitude of the superconducting order parameter within the puddle, and consequently $\xi_0$ itself, depends sensitively on *l*.

Our results are collected together in Fig. S9. In Fig. S9a, we compare the spectral gap (here rather roughly defined as half the peak-to-peak distance in energy, measured from the LDOS) to the *d*-wave order parameter, with both quantities averaged within the puddle. It can be seen that the spectral gap tracks the *d*-wave order parameter closely for the bulk system and for larger puddles (7 × 7 up to 17 × 17, all corresponding to 2 < $l/\xi_0$ < 10), although for puddles the spectral gap slightly *overestimates* the *d*-wave order parameter. However, once the puddle size is small enough such that $l/\xi_0 \approx 1$ (as is the case for the 5 × 5 puddle), a gap is no longer visible in the spectrum, even though a nonzero superconducting order parameter remains within the puddle. The strong pair-breaking effects of the surrounding metal are most easily seen in Fig. S9b. Here we plot the average *d*-wave order parameter within the puddle as a function of $l/\xi_0$ The average order parameter within all seven puddles considered is considerably less than that of the bulk system, and decreases in magnitude as $l/\xi_0$ is made smaller. Note that when the puddle is made smaller and smaller, the mixing of metallic states into the superconducting puddle increases since more of the puddle becomes in closer proximity with the superconductor-metal boundary, and hence there is more pair-breaking. Fig. S9c shows the average zero-energy LDOS for seven different puddle sizes. Notice that the zero-energy LDOS of all puddles is much bigger than that of the bulk system, and that it increases as the puddle size is decreased. As with the *d*-wave order parameter, the large zero-energy LDOS is an effect of the mixing of the metallic states into the superconducting puddle, giving the latter a much larger number of low-energy states than one would expect a bulk *d*-wave superconductor to have. The overall trend is succinctly captured by Fig. S9d, which plots together with the *d*-wave order parameter and the zero-energy LDOS both averaged within the puddle, with the variations in both quantities due solely to the puddle size. It can be seen that these two quantities are inversely proportional to each other, with a large *d*-wave order parameter corresponding to a small zero-energy LDOS and vice versa. This is behavior very similar to that expected from disorder acting on a bulk *d*-wave superconductor; one cannot escape the conclusion that the metallic matrix induces pair-breaking effects within the superconducting puddle very similar to that of disorder.

All of these findings are more explicitly demonstrated in Fig. S10, wherein we show plots of the LDOS vs. energy for three puddle sizes (5 × 5, 7 × 7, and 9 × 9, whose $l/\xi_0$ values are given approximately by 1.2, 2.4, and 3.6, respectively), in addition to the bulk *d*-wave case. We note first that for superconducting puddles, a striking feature of the LDOS is its very large value at $E = 0$ compared to that of the bulk system. One can also notice that for larger puddles, a gap is easily discerned in the spectrum, and coherence peaks are visible but are broader, less well-defined, and shorter in height compared to those of a bulk system. These features become progressively broader as the puddle is shrunk, and more spectral weight accumulates near the Fermi energy, a result of the fact that the average *d*-wave order parameter becomes smaller the tinier the puddles get. However, when the puddle is made sufficiently small such that $l/\xi_0 \approx 1$, such as the 5 × 5 case here, the gap ceases to be visible in the quasiparticle spectrum, and the LDOS resembles that of a normal metal. Nevertheless, there is still a nonzero *d*-wave order parameter present within the puddle.

In sum, we have shown here some of the surprising effects of embedding a *d*-wave superconducting puddle within a metallic matrix. We have demonstrated that the surrounding metallic matrix has a pair-breaking effect on the superconductivity within the puddle, akin to that of disorder, that fills the gap, including at the Fermi level. We have also shown that the smallness of the puddle has a nontrivial effect on the LDOS, with the quasiparticle spectrum within the puddle showing broad signatures of a gap that progressively becomes filled up and washed out the smaller the puddle becomes. The similarity of the pair-breaking effects of the metallic matrix to disorder points to the difficulty of attributing the effects seen in the experiment and detailed in the main text to purely mean-field effects.

Our calculations show that within a mean-field picture, pair-breaking, whether it be due to disorder or the effect of metallic states on a superconducting puddle, naturally leads to an *anticorrelation* between these two quantities. Such a scenario points to the necessity of "beyond-mean-field" physics in resolving the conundrum posed by the experimental results discussed in the main text.

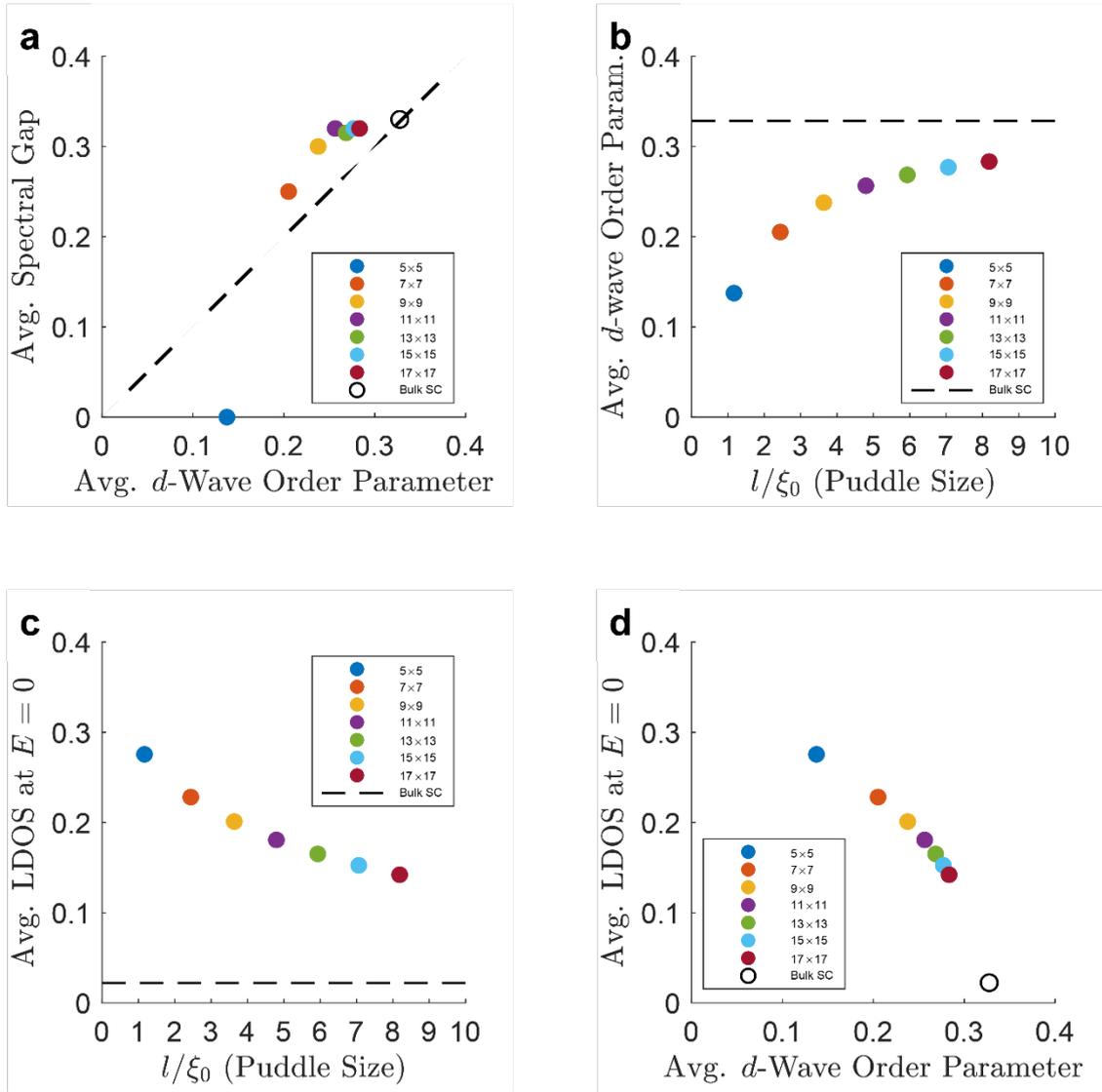

**Figure S9:** Results for clean superconducting puddles within a metallic matrix of varying size, with $V_0$ = 1.0. A (top left): plot of the average spectral gap versus the average $d$-wave order parameter, both averaged within the superconducting puddle, with the dashed line indicating where the two quantities are equal. It can be seen that for puddles, the spectral gap is a good indicator of the $d$-wave order parameter (although overestimating it, lying above the dashed line) right until the puddle becomes sufficiently small and $l/\xi_0 \approx 1$ (*e.g.*, 5 × 5), at which point no gap can be seen even though a nonzero superconducting order parameter is present. B (top right): plot of the average of the $d$-wave order parameter within the superconducting puddle versus the puddle size $l/\xi_0$, with the value for the bulk system shown as a dashed line. C (bottom left): plot of the zero-energy LDOS averaged within the superconducting puddle versus the puddle size $l/\xi_0$, again with the value for the bulk system shown as a dashed line. D (bottom right): plot of the LDOS at $E$ = 0 versus the $d$-wave order parameter, both averaged inside the superconducting puddle. Evidently, the effect of reducing the puddle size on the superconducting condensate within the puddle is the same as that of increasing the amount of disorder: the $d$-wave order parameter becomes smaller, while the zero-energy LDOS increases.

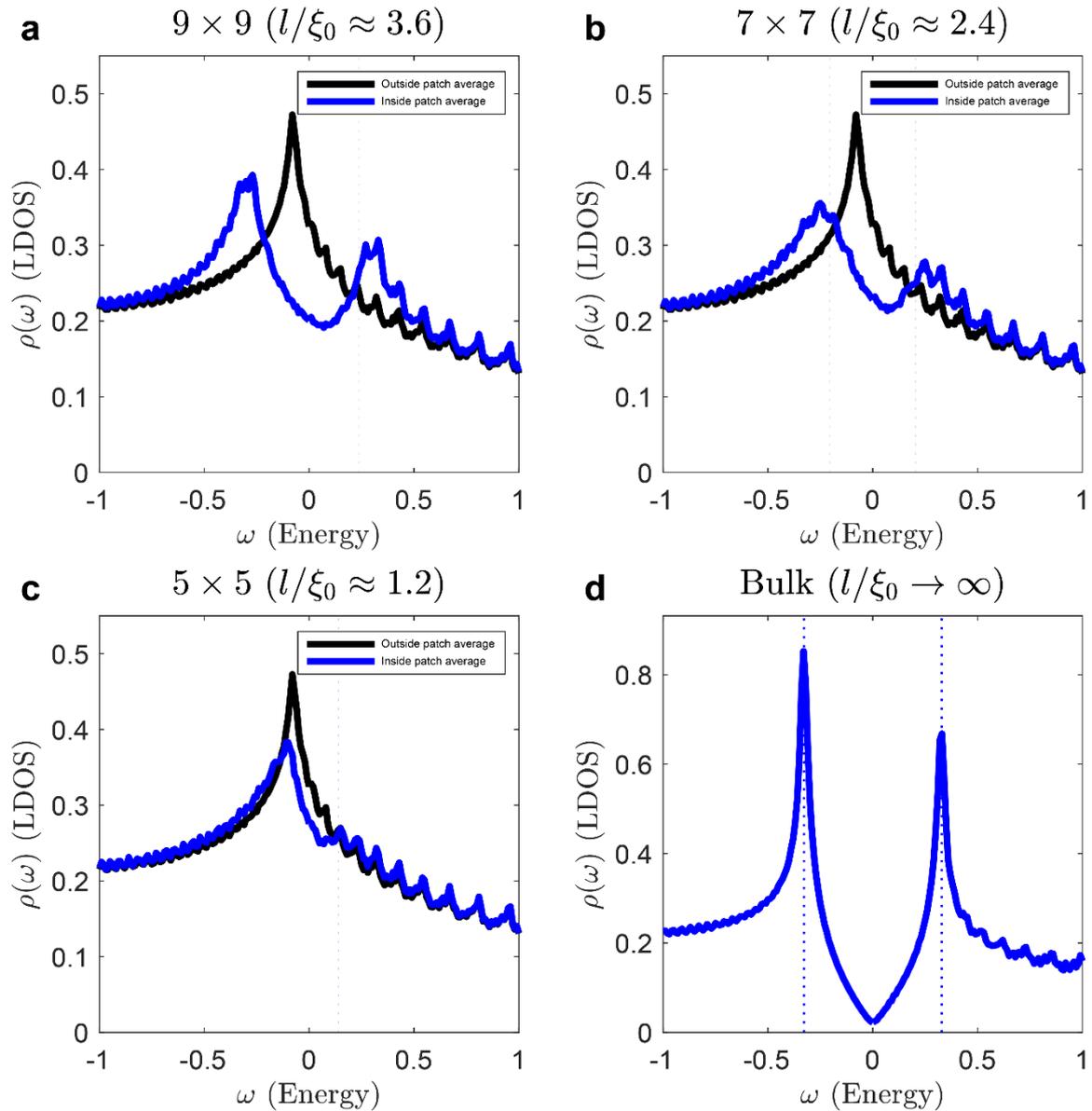

**Figure S10:** Plots of the LDOS as a function of energy for a *d*-wave superconducting puddle inside a metallic matrix with decreasing puddle size, with $V_0$ = 1.0. The puddle sizes are 9 × 9, 7 × 7, and 5 × 5 (A-C), corresponding to $l/\xi_0$ approximately equal to 3.6, 2.4, and 1.2, respectively. Shown are LDOS averages within the puddle (blue) and outside it (black). Also shown for comparison is the average LDOS for a bulk *d*-wave superconductor with the same pairing interaction $V_0$ = 1.0 (D). The dashed blue lines indicate the average *d*-wave order parameter within the superconducting puddle. Note that as the puddle size becomes smaller, the zero-energy LDOS inside the patch becomes larger, the coherence peaks become smeared out and move to lower energies, and the gap becomes less discernible.

## V. Rigid band shift in overdoped Bi2201

ARPES measurements on SOD samples show a rigid band shift of the anti-nodal band bottom when the doping is increased (Figure S11). ARPES measurements were performed using the He1α line at 21.2 eV with linear polarization. The sample temperature for all samples was 50K, and the total experimental resolution was set to 6 meV. The k-space cut was along the face of the Brillouin zone (a line spanning the direction (π,π)-(π,0)-(π,-π), indicated in the inset of Fig S11a).

Shown in Fig. S11a as "+" symbols are the positions of the peak maxima of the Energy Distribution Curves (EDC's), extracted after dividing out the resolution broadened Fermi-Dirac distribution from the raw data. The energy position of the band bottom as shown in Fig. S11b is extracted by taking the average of the EDC maxima positions within a small momentum window (indicated in Fig S11a by the two vertical red lines). Using these doping-dependent band bottom energy values, we can determine the doping level of the OD0K crystal by fitting the positions of the superconducting samples and extrapolating the result. We find that the OD0K sample has a doping level of *p*=0.274±0.008.

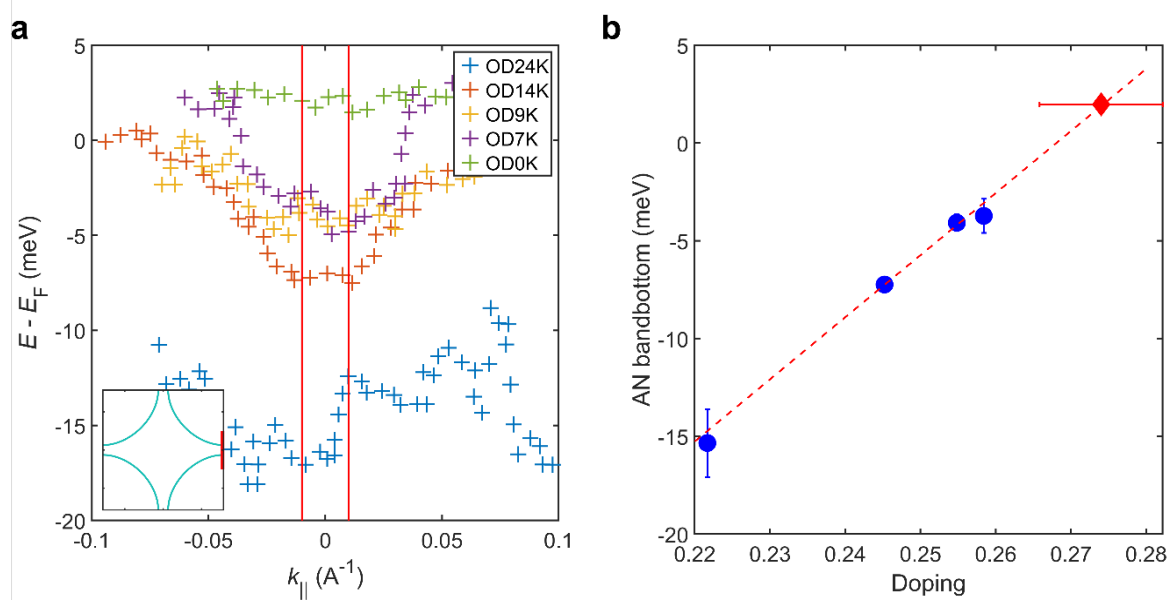

**Figure S11:** Rigid band shift of the anti-nodal band bottom. The anti-nodal ARPES cut is indicated by the red line on the Fermi surface in the inset to panel (a). In (a) can be seen that the anti-nodal band bottom shifts towards the Fermi level as the samples are progressively overdoped. The energy position of the band bottom is shown in panel (b), and is given by the average position within the red vertical lines in panel (a). The red dashed line in (b) shows a fit through the band bottom energy position for the superconducting samples (blue circles). The fit is then extrapolated to determine the doping level of the non-superconducting sample (red diamond), given its measured band bottom.